# Internationalizing AI: Evolution and Impact of Distance Factors

Xuli tang (0000-0002-1656-3014)

School of Information Management, Central China Normal University, Wuhan 430079, China

Xin Li * (0000-0002-8169-6059)

School of Medicine and Health Management, Tongji Medical College, Huazhong University of Science and Technology, Wuhan 430030, China

Feicheng Ma (0000-0003-0187-0131)

School of Information Management, Wuhan University, Wuhan 430074, China

* Corresponding concerning this article should be addressed to Xin Li (xint@hust.edu.cn). The address of corresponding author is School of Medicine and Health Management, Tongji Medical College, Huazhong University of Science and Technology, Wuhan 430030, China.

**Abstract**

International collaboration has become imperative in the field of AI. However, few studies exist concerning how distance factors have affected the international collaboration in AI research. In this study, we investigate this problem by using 1,294,644 AI related collaborative papers harvested from the Microsoft Academic Graph (MAG) dataset. A framework including 13 indicators to quantify the distance factors between countries from 5 perspectives (i.e., geographic distance, economic distance, cultural distance, academic distance, and industrial distance) is proposed. The relationships were conducted by the methods of descriptive analysis and regression analysis. The results show that international collaboration in the field of AI today is not prevalent (only 15.7%). All the separations in international collaborations have increased over years, except for the cultural distance in masculinity/felinity dimension and the industrial distance. The geographic distance, economic distance and academic distances have shown significantly negative relationships with the degree of international collaborations in the field of AI. The industrial distance has a significant positive relationship with the degree of international collaboration in the field of AI. Also, the results demonstrate that the participation of the United States and China have promoted the international collaboration in the field of AI. This study provides a comprehensive understanding of internationalizing AI research in geographic, economic, cultural, academic, and industrial aspects.

## 1. Introduction

Just as Adams declared, international collaboration has driven us into a fourth era of research with best science (Adams, 2013); It has become imperative in various research fields (Coccia & Wang, 2016). Ribeiro et al. (2018) argued that international collaboration has been increasing faster than an exponential growth, based on the analysis of over 10 million articles indexed in the Web of Science (Ribeiro et al., 2018). International collaborations bring together researchers from different countries working for common goals by sharing knowledge, resources, and experience, which can enhance academic impact, inspire novel ideas, and carve out successful career paths for scholars (Wagner et al., 2019; Yao et al., 2020; C. Zhang et al., 2018). For example, Wagner et al. (2017) demonstrated that the number of collaboration countries per paper has significant positive correlation with higher citation counts by examining international collaboration in six traditional disciplines (i.e., engineering, mathematic, astrophysics, virology, agriculture, and seismology).

Evidences have also indicated that international collaboration is critical to the AI community for serving humans in more complicated situations. For examples, with the global outbreak of COVID-19, AI scientists have internationally collaborated with domain experts, such as virologists, clinicians, and epidemiologists (Bullock et al., 2020), on automatic image diagnosis (Ai et al., 2020), drug repurposing (Hoffmann et al., 2020), and global epidemic prediction (Al-qaness et al., 2020). On the one hand, these collaborations promote the sharing of COVID-19 related research resources, such as clinical cases, scientific literature, and datasets globally. On the other hand, they not only help AI scientists deeply understand the outputs of their AI algorithms, but also provide domain experts with novel insights on virus. International collaboration in AI has been speeding up the process of combating the COVID-19. Despite the significant benefits and the prosperity in traditional disciplines, it is worth to note that international collaboration is still not prevalent in the field of artificial intelligence (Niu et al., 2016; Tang et al., 2020; Tang et al., 2021). This conflict raised a question: What factors affect the international collaboration in AI?

Much of recent studies have endeavored to identify the determinants affecting international collaboration in several research fields, which can be summarized into four categories, including geographic distance (Abramo et al., 2020; He et al., 2020; Hoekman et al., 2010; Parreira et al., 2017; Sidone et al., 2017; Yao et al., 2020), economic distance (Acosta et al., 2011; Fernández et al., 2016; He et al., 2021a; He et al., 2021b; Jiang et al., 2018; Ni & An, 2018), cultural distance (Cassi et al., 2015; Choi et al., 2015; Gervedink Nijhuis et al., 2012; Gui et al., 2019; Hoekman et al., 2010; Hung, 2008; Jiang et al., 2018; Plotnikova & Rake, 2014), and academic distance (Fernández et al., 2016; Kwiek, 2015; Osiek et al., 2009; Parreira et al., 2017). These determinants have been together or individually examined in limited disciplines, and their performance was different because of the underlying differences in disciplines. In the field of marketing, geographic distance, and cultural distance both have no significant relevance with international collaborative publications (Jiang et al., 2018). But in the field of ecology, geographic distance academic distance (the distance of scientific production) and other socioeconomic factors (e.g., human development index) explained around 10% of the international collaboration. However, there is no studies on how these factors influence the international collaboration in the field of AI. Meanwhile, AI is intrinsically a discipline that is highly relevant to the industry. On the one hand, evidence has shown how fast the new emerging AI algorithms have been applied in the industry. On the other hand, after the late of 1980s, industry-based research affiliations, such as Google and Microsoft, are increasingly becoming the central of the modern AI research (Frank, Wang, et al., 2019). Therefore, how the difference of industrial participation in AI research between countries (i.e., industrial distance) affects the international collaboration in AI should also be explored.

Understanding the relationships between these distance factors and the international collaboration in the field of AI can provide a deep insight on how to achieve successful and efficient international collaboration in AI research. To examine these relationships, in this paper, we harvested 1,294,644 AI collaborative papers between 1950 and 2019 from the Microsoft Academic Graph (MAG). We combined the methods of descriptive analysis and regression analysis. Specifically, we first analyzed the overview of international collaboration in the field of AI, as well as the trends of the geographic, economic, cultural, academic, and industrial separations of international collaborations in the field of AI over years. Then we examine the relationships between these distance factors and international collaboration in AI using the zero inflated beta regression model and multiple linear regression model.

The results of our analysis show that international collaboration in the field of AI today is not prevalent (only 15.7 of collaborative paper are international collaborations). The geographic distance, economic distance, and academic distances between countries have shown significantly negative relationships with the degree of international collaborations in the field of AI. Meanwhile, the industrial distance has a positive relationship with the international collaboration in the field of AI. Finally, the results demonstrate that the participation of the United States and China have promoted the international collaboration in the field of AI.

## 2. Related work

### 2.1. International collaboration and its advantages in different disciplines

Overall, the recent increase of international collaboration in science has surpassed the exponential growth (Ribeiro et al., 2018; Wagner et al., 2019), however, it progresses unevenly in different disciplines. For example, it has been proven that the virology is the most internationalized discipline with 120 countries involved in international collaboration in 2013, while the mathematics is the least internationalized one with just 35 countries taking part in international collaboration in 2013 (Wagner et al., 2017). Natural science (22.7% in 2011) has a slightly higher degree of internationalization than social sciences (16.4% in 2011) (Larivière et al., 2015).

International works are widely believed to have significant advantages over works co-authored by researchers within a nation (Jiang et al., 2018; Kwiek, 2015), because of the impetus of international collaboration in promoting the information exchange, enhancing the team diversity, attracting more citations, improving research quality and inspiring novel ideas (Adams, 2013; Coccia & Wang, 2016; Larivière et al., 2015; Moaniba et al., 2019; Wagner et al., 2017, 2019). For instance, Larivière et al. (2015) analyzed billions of papers and their citations in both natural science and social science between 1900 and 2011. They found that the larger the number of countries per article, the higher the impact, which cannot be interpreted by self-citations. Adams (2013) investigated a large number of academic papers in various countries and discovered that the internationally collaborative papers had attracted more citations than the domestic papers in all disciplines. Moaniba et al. (2019) utilized the granger causality test on four million of patents in USPTO and macroeconomic data for 54 countries from 1976 to 2015 and revealed that positive relationship between international collaboration and the intensity of innovation. This result was robust across groups of countries and different time periods.

Considering these significant advantages, international collaboration has been imperative in various research domains, especially for the emerging domains such as the artificial intelligence (AI). In the past half century, AI has experienced the periods of the sprout, the three waves and the rapid development. Nowadays, it has drastically changed the way of our work and life. For instance, the marketization of automation technologies, such as autonomous car, natural language question-answering, smart finance, and precision medicine, have its potential in reshaping skill demands, reducing career opportunities and disrupting labor markets (Frank, Autor, et al., 2019). AI is a fast applied but unmatured area, and the lifting of the usability and interpretability of the results apparently required the extensive collaboration between AI scientists and domain experts (Pinhanez, 2019). Inter-institutional collaboration has been assumed as the strategy AI advances itself (Shao et al., 2020). After analyzing all the papers published by the top 2000 high-cited AI scholars during last decade (2009-2019) based on the AMiner, Shao et al. (2020) found that international collaboration in the papers of top AI scientisits occupied more than half of inter-institution collaboration. However, despite the significant advances in AI (Frank, Wang, et al., 2019), international collaborations are still not prevalent in the whole field of AI (Niu et al., 2016). Therefore, it is crucial to explore which measurable factors impede the international collaboration in AI. To our best knowledge, few studies have attempted to systematically reveal the relationships between external factors and international collaboration in the field of AI.

## 2.2. Factors affecting international collaboration

International collaboration is not easy to achieve because scholars need to overcome various obstacles that may impede the communication and collaboration (Yao et al., 2020). Previous studies have provided some insights on these factors, which can be categorized into four aspects: geographic distance, economic distance, cultural distance, and academic distance.

The rapid development of information and communication technology (ICT) has made us think that geographic distance would no-longer hinder the international collaboration, because the online interactions that bridge between authors, provide more collaboration opportunities, and facilitate knowledge flow (Ding et al., 2010), become non-costly as physical distance increases (Yao et al., 2020). However, it seems that the function of ICT is overestimated. After analyzing copublications between European countries, Hoekman et al. (2010) demonstrated that the european integration and the development of ICT cannot change the fact that collaboration in European is still sensitive to the geographic distance (Hoekman et al., 2010). This result is consistent with Parreira et al. (2017) and Sidone et al. (2017), who found that international collaboration among scholars decreases as the geographic distances among countries increase (Parreira et al., 2017; Sidone et al., 2017). Besides, it was discovered that the geographic distance was a dominant fator affecting the knowledge flow in collaboration networks (Abramo et al., 2020; Sidone et al., 2017), and the role of geographic distance on knowledge flow differs across disciplines: it has a significant effect on agricultural science, engineering, and health sciences but limit effect on humanities and social sciences (Abramo et al., 2020).

Economic distance has been highlighted as an explanatory factor for international collaboration (Fernández et al., 2016). The economic level of a country usually represents the occupation of resources and the opportunities providing for mobility and participation in international conferences. Based on the analysis of scientific collaborations among European countrie, Acosta et al. (2011) found that countries that invested similar level of resources in R&D were more inclined to collaborated with each other (Acosta et al., 2011). Similarity, Ni and An (2018) analyzed the papers in the field of Public, Environmental and Occupational Health, and demonstrated that the largest percentage of international works are co-published by countries with similar economic level (Ni & An, 2018). Jiang et al. (2018) found that economic distance was negatively correlated to the number of international works for countries, then they referred that countries with similar level of economics accounted for the majority of international works (Jiang et al., 2018).

Cultural distance refers to the difference in code of conduct and communication, and it determines the success in establishing mutual trust and mutual respect among international collaborators (Plotnikova & Rake, 2014). The culture proximity has been broadly studied from both linguistic and social levels. At the linguistic level, a common language is beneficial to international collaboration for it enables scholars to better exchange their ideas (Yow & Lim, 2019) and understand culturally meanings embedded (Jiang et al., 2018) during their communications. Gui et al. (2018) analyzed a large number of articles from the Web of Science and demonstrated that a common language significantly promoted international collaboration. However, this is inconsistent with the studies of Parreira et al.(2017) and Cassi et al.(2015), in which they found that English as official language had insignificant effect on international collaboration ( Cassi et al., 2015; Parreira et al., 2017).

For cultural distance at the social level, Hofstede (2001) proposed six kinds of dimensions from the theoretical perspective: power distance, uncertainty avoidance, individualism/collectivism, masculinity/femininity, long-term orientation, and indulgence/restraint (Bhagat & Hofstede, 2001; Hofstede, 2010), which provide a theoretical base for quantifying cultural distance at social level. For example, Gervedink Nijhuis et al. (2012) demonstrated that power distance and individualism/collectivism determined the communication and conflict resolving among collaborators.

Hung (2008) found that team member's individualism/collectivism have significant impact on their levels and their forms of participation. By contrast, after combining culture dimensions as one index, Jiang et al. (2018) found it has no significant associations with the international collaboration.

Academic distance reflects the difference in the academic strength between collaboration pairs. Previous studies have mainly examined academic distance in the context of scientific collaboration from two perspectives: the academic production and the academic influence. The distance in academic production between countries has been widely measured by the difference of the number of scientific publications (Fernández et al., 2016; Kwiek, 2015), and the distance in academic influence has been always represented by the difference of the number of citation counts between countries (Parreira et al., 2017). As Fernández et al. (2016) indicated, the number of collaborations increases as the number of publications of each collaborators rises, that is, collaboration has a correlation with academic production of collaborators. Kwiek (2015) demonstrated that the degree of international collaboration in Europe is significantly related to the number of publications published by countries. For the aspect of academic influence, Parreira et al. (2017) calculated the Euclidean distance between countries based on the total number of citation counts of all the papers and revealed that the distance in academic influence has a significantly negative relationship with international collaboration.

Industries has been playing significant role in the scientific research, especially for the leading-edge domains such as artificial intelligence (AI). On the one hand, industrial research affiliations, such as Google, Bell Lab, Microsoft, have always been in a leading position in AI research (Frank, Wang, et al., 2019). On the other hand, AI is the key enabler for the transformation of the Science andTechnology outputs. Ahmadpoor and Jones (2017) found that the average links of a AI patent backward to a basic research paper is 1.5, while the average degree for computer science is 2.35 and the average degree for mathematics is around 5 (Ahmadpoor & Jones, 2017). AI has gradually disrupted the old value chain and transformed entire industries, from music (Hujran et al., 2020) to insurance (Sushant K, 2020), banking (Shie et al., 2012), publishing (A. Huang, 2019) and transportation (Nikitas et al., 2020). Thus, in this paper, we propose industrial distance as an index to measure the difference of the degree of industrial participation between countries and explore how it influences the international collaboration in AI.

## 3. Methods

To reveal the effects of distance factors on international collaboration in the discipline of AI, we first introduce the detail of data collection and pre-processing. Second, we use 13 indicators to quantify the distance factors between countries from 5 perspectives (i.e., geographic distance, economic distance, cultural distance, academic distance and industrial distance).

We also use the Jaccard coefficient to measure the degree of international collaboration in AI between countries. Based on these indicators, we further characterize the trends of international collaboration in the field of AI over years, and finally explore how these distance factors affect the international collaboration in the field of AI. The overview of the research is illustrated in Figure 1. Detailed information on every step is explained as follows.

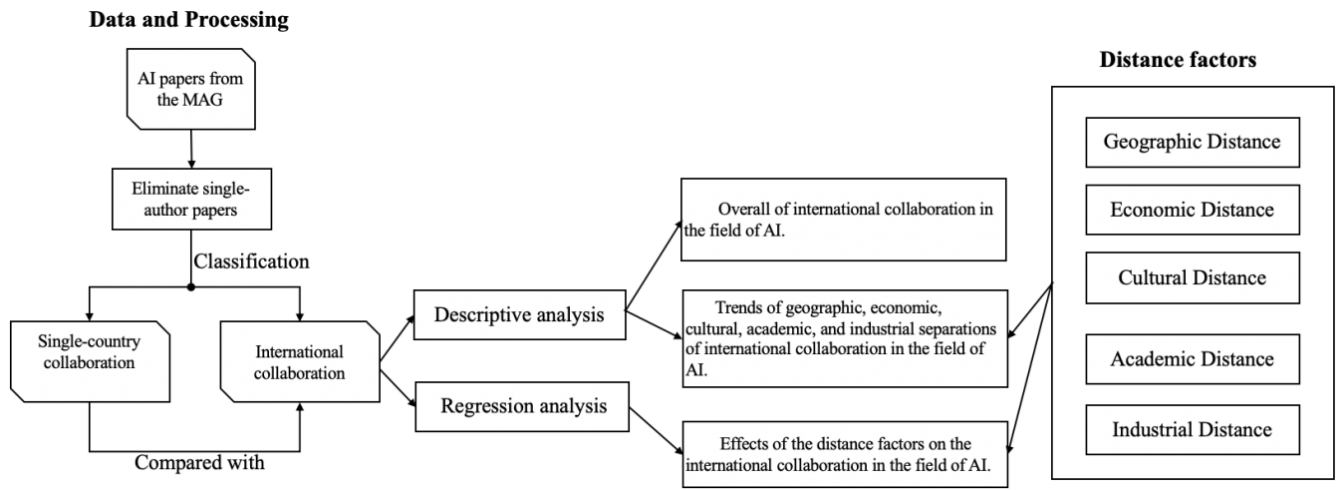

Figure 1. Overview of this study.

## 3.1. Data and processing

The dataset used in this paper is derived from the Microsoft Academic Graph (MAG), which is a heterogeneous graph containing multiple types of publications (e.g., journal papers, conference papers, patent, book, book chapters or other) and the citation relationships between those publications. The MAG dataset has been widely used and evaluated in previous scientometric research (Frank et al., 2019; Huang et al., 2020). In this study, we used publications in the subfields of artificial intelligence, machine learning, computer vision, nature language processing and pattern recognition to represent AI publications (Frank et al., 2019). We download the data from Microsoft Azure platform and deleted records whose document type is not journal paper, conference paper or patent. Bibliographic information of each AI paper such as title, year, abstract, author, and affiliations with its latitude and longitude were extracted and stored in a local MySQL database. The country for each affiliation was recognized using its latitude and longitude. Then, single-authored papers and those papers without metadata were eliminated. The final dataset for analysis includes 1,294,644 AI papers published from 1950 to 2019, in which the number of international collaborated papers was 202,799 and the number of papers with a single country was 1,091,845.

Each paper in AI was assigned to one or more countries according to the nationality of the authors' affiliations, and there was total 169 countries were identified. To calculating the distance factors, for each country, we obtained the following information: (1) the geographic coordinates; (2) the average gross domestic product (GDP) per capita in recent ten years obtained from the world bank (https://data.worldbank.org/indicator/NY.GDP.PCAP.CD); (3) the Hofstede's six culture indexes (including power, uncertainty avoidance, individualism/collectivism, masculinity/femininity, long-term orientation and indulgence/restraint) from Geert Hofstede's website (https://geerthofstede.com/research-and-vsm/vsm-2013) (Hofstede et al., 2010); (4) Whether English is its primary or official language; (5) the total number of AI papers and the number of internationally collaborative AI papers; (6) the number of citations received by the International collaborations in the field of AI from the MAG dataset; (7) the number of academic conferences held by the country from the file ConferenceInstances.txt in the MAG, which includes the location (latitude and longitude) for 16,083 conferences worldwide; (8) the number of industry-involved AI paper. We marked an affiliation as industrial if it is an instance of "company", "business" or "enterprise" in the WikiData (https://www.wikidata.org/).

*3.2. Measuring the distance factors*

The framework for analyzing the relationships between distance factors and the international collaborations between countries in the field of AI is illustrated in Figure 2. In this section, we give a detailed description on how to calculate the thirteen distance factors from five different perspectives.

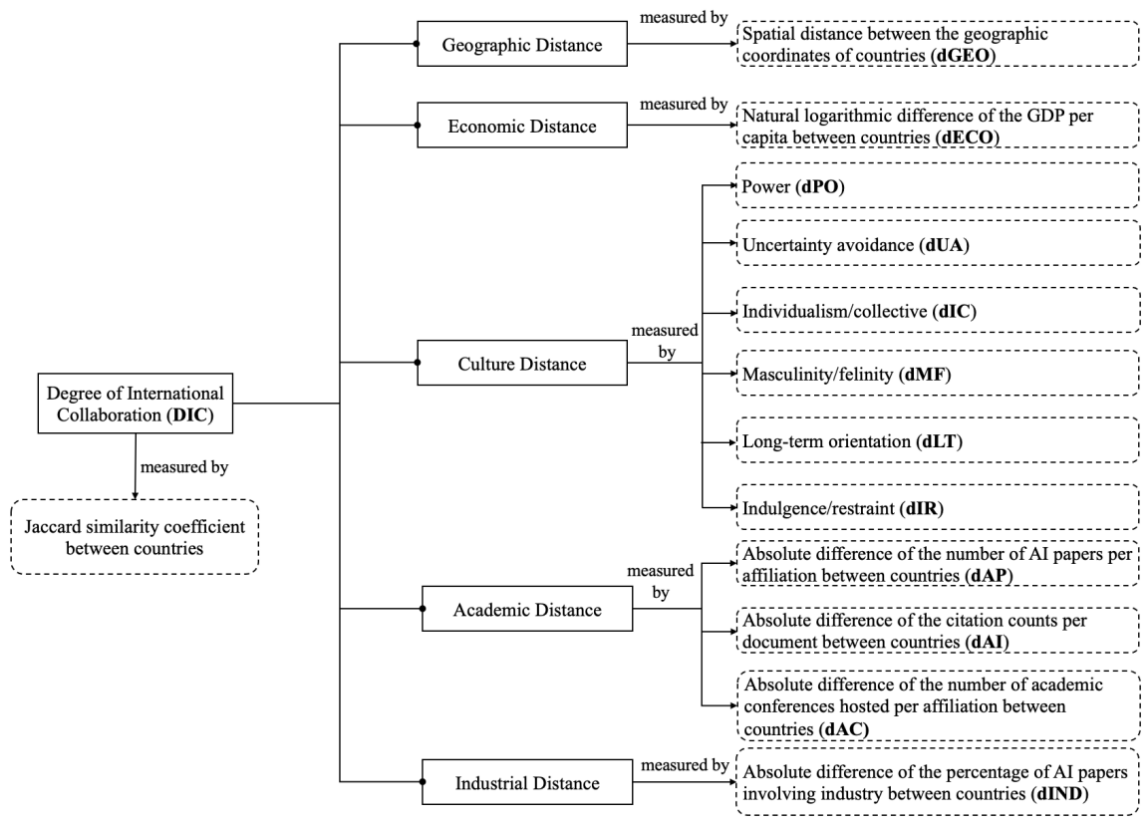

Figure 2. The framework for analyzing the relationships between distance factors and the international collaborations between countries in the field of AI.

**Geographic distance (dGEO)** is defined as the kilometers between a country pair. Prior studies mainly used the kilometers(km) between the capital cites as proxy of the geographic distance between countries (Gui et al., 2018, 2019; Jiang et al., 2018; Parreira et al., 2017). However, this method may lead to an imprecise result as the number of publications of different research affiliations is uneven in a country. Instead, in this paper, the geographic coordinates for each country were evaluated by the center of gravity of the polygon which is comprised of the longitudes and latitudes of research affiliations. Thus, for a given country pair (i, j), we use the great circle formula (Su & Moaniba, 2020) to calculate its geographic distance, as expressed by:

$$dGEO = r \times \arccos\left[\sin\left(\frac{lat_i \times \pi}{180}\right) \times \sin\left(\frac{lat_j \times \pi}{180}\right) + \cos\left(\frac{lat_i \times \pi}{180}\right) \times \cos\left(\frac{lat_j \times \pi}{180}\right) \times \cos\left(\frac{lon_i \times \pi}{180} - \frac{lon_j \times \pi}{180}\right)\right], \quad (1)$$

where r represents a constant value of the radius of Earth (r=6377km); $lat_i$ and $lon_i$ indicate the latitude and longitude of the country $i$, $lat_j$ and $lon_j$ indicates the latitude and longitude of the country $j$.

**Economic distance (dECO)** of a specific county pair is measured by the natural logarithmic difference of gross domestic product (GDP) per capita between each country pair. The GDP per capita (current US dollar) is broadly accepted as a basic evaluation of economic development level of a country. However, the difference between countries' GDP per capita is too far from each other. In econometrics, the natural logarithmic difference is commonly adopted for the normalization (Jiang et al., 2018; Tsang & Yip, 2007). Thus, the economic distance of a given country pair (i, j) is given by:

$$dECO = |\lg(GDP_i) - \lg(GDP_j)|, \quad (2)$$

where $GDP_i$ and $GDP_j$ indicate the GDP per capita for the country $i$ and $j$, respectively. We also use the absolute value of the difference to ensure it is positive.

**Cultural distance** measures the culture differences between countries from different cultural dimensions. In this paper, we apply the commonly adopted metrics - the Hofstede's six culture dimensions - to quantify the difference in cultural aspects between countries. The Hofstede's six culture dimensions are comprised of the power (**PO**), the uncertainty avoidance (**UA**), the individualism/collectivism (**IC**), the masculinity/felinity (**MF**), the long-term orientation (**LT**), and the indulgence/restraint (**IR**) (Cashman et al., 2019; S. (Sam) Huang & Crotts, 2019; Jiang et al., 2018). In this paper, we use the absolute difference of the cultural indicators to quantify the cultural difference between countries. For example, for a given country pair (i, j), the culture distance in the power dimension (**dPO**) is calculated by:

$$dPO = |PO_i - PO_j|, \quad (3)$$

where the $PO_i$ and $PO_j$ represent the power index of the country i and j that were derived from Hofstede's website. The cultural distance in other five dimensions between countries (i.e., **dUA, dIC, dMF, dLT** and **dIR**) can be easily calculated using the same method.

**Academic distance** of a given country pair, in this paper, is measured from three perspectives, including academic production, academic influence and academic communication. For calculating the academic distance in academic production (**dAP**), we use the absolute difference of the number of AI papers per affiliation between the two countries. We then quantify the academic distance in academic influence (**dAI**) between countries by using the Euclidean distance between countries based in their total number of AI citations per document. As a formal academic communication format, academic conferences promote the collaboration among scholars worldwide. Therefore, for measuring the academic distance in academic communication (**dAC**), in this paper, we use the absolute difference of the number of academic conferences hosted per affiliation between the two countries.

**Industrial distance (dIND)** of a given country pair is defined as the difference of the industry involvement in the research of AI between the two countries. Specifically, we use the absolute difference of the percentage of industry-involved AI papers between the two countries to measure this distance. Then, the dIND of a country pair (i, j) is calculated by:

$$dIND = |P_{IND_i} - P_{IND_j}|, \quad (4)$$

where $P_{IND_i}$ and $P_{IND_j}$ indicate the percentage of papers having at least one industrial affiliation for the country i and j, respectively.

We also consider the influence of English as primary language or official language on international collaboration in the field of AI. **English as primary language or official language (ENG)** is a dummy variable, for a given country pair, if both countries have the English language as their primary or official language, then ENG = 2; else if one of the countries have English as its primary or official language, then ENG = 1; or ENG = 0 (Gui et al., 2018; Parreira et al., 2017). In addition, according to our analysis, we found that the United States and China have played important roles in the international collaborations in the field of AI, hence, we consider the participation of these two countries. Specifically, the dummy variable **CoUS = 1** (or **CoCN =1**) if one of the country pair is the United States (or China); or **CoUS = 0** (or **CoCN =0**).

### 3.3. Measuring the degree of international collaboration between countries

To quantify the strength of collaboration ties between country pairs, in the light of previous studies (Gui et al., 2018, 2019; Parreira et al., 2017; Shi et al., 2020), we here adopt the Jaccard similarity coefficient, which can capture the degree of overlap of two countries' scientific research in the AI discipline. Thus, for a given country pair (i, j), **the degree of international collaboration (DIC)** in AI between them is calculated by:

$$DIC_{ij} = \frac{C_{ij}}{C_i + C_j - C_{ij}}, DIC_{ij} \in [0,1] \quad (5)$$

where $C_i$ and $C_j$ indicate the number of AI papers produced by the country $i$ and country $j$, respectively; and $C_{ij}$ represents the number of AI papers jointly collaborated by the two countries. A larger the Jaccard similarity coefficient indicates a stronger degree of collaboration tie between the two countries. The value of DIC ranges from 0 and 1; and DIC = 1 represents that all AI papers produced by the country pair were in collaboration with each other, while DIC = 0 represents none of the AI papers were collaborated by the country pair.

In addition, we also analyze the international collaboration in the field of AI using the method of social network analysis. We construct a collaboration network in AI among all countries in our dataset, whose nodes are countries and edges represent the collaboration relationships between countries. We quantify the importance of a country in AI collaboration by using the **relative degree centrality (RDC),** which was employed to assess the significance of countries in collaborative networks with different scales (G. Zhang et al., 2014)**.** Specifically, for a given county i, the relative degree centrality of it can be represented by:

$$RDC_i = \frac{DC_i}{N-1} = \frac{\sum_{j=1}^{N-1} e_{ij}}{N-1}, \quad (6)$$

where $e_{ij}$ is 0 or 1, and $e_{ij}$ =1 indicates there is an edge between country $i$ and country $j$, vice versa. The $DC_i$ means the number of connections of the country i, and (N -1) represent the total number of countries (nodes) in the network, which is used for normalization (X. Zhang & Chai, 2019).

### 3.4. Regression analysis

To explore the potential relationships between international collaboration and the five types of distance factors in the field of AI, we model the degree of international collaboration (DIC) between countries as the response variable using two regression analysis. Considering that the values of DIC range from 0 to 1, and more than half of them are 0, we choose the model of zero inflated beta regression (ZIBeta) (Ospina & Ferrari, 2012; Parreira et al., 2017). Besides, we compare our results with a standard multiple regression model (i.e., Ordinary Least Squares, OLS), which is the most used regression model in research fields, such as econometrics and scientometrics. All the explanatory variables are rescaled between 0 and 1 using min-max normalization before the regression to make partial coefficients comparable to each other.

## 4. Results

### 4.1. Overview of International Collaboration in the Field of Artificial Intelligence

International collaboration is still not prevalent in the AI discipline. There is total 1,294,644 collaborative papers on AI in the MAG dataset from 1950 to 2019, and only 15.7 % of which (202,799) are internationally collaborated. Although national collaboration is still dominant in the field of AI, the percentage of internationally collaborated AI papers has clearly increased over years (Figure 3), with an increasing growth rate. For example, in 1978, the ratio of international collaboration in AI was 4.8%, and it took 28 years to double this ratio; whereas it only cost 14 years to increase from 11.3% in 2006 to 22.8% in 2019. On the contrary, the percentage of single-country collaborated AI papers has shown a clear decline since 1980. Meanwhile, the annual percentage of citations received by international collaborations in AI has been increasing over years. Specifically, before 1990, the annual percentage of citations received by international collaborations has kept the same level with that of the internationally collaborated AI papers with fluctuations. Then, it started to surpass that of the internationally collaborated AI papers and the gap between them is widening over time. In 2019, 36.3% of the citations were contributed by the international collaborations (accounting for around 21%) in AI. Besides, we can also observe obvious fluctuations of the citations in the year 1954 and 1973, which may be caused by the small number of AI papers and several multi-country collaborated AI papers that has high citation counts.

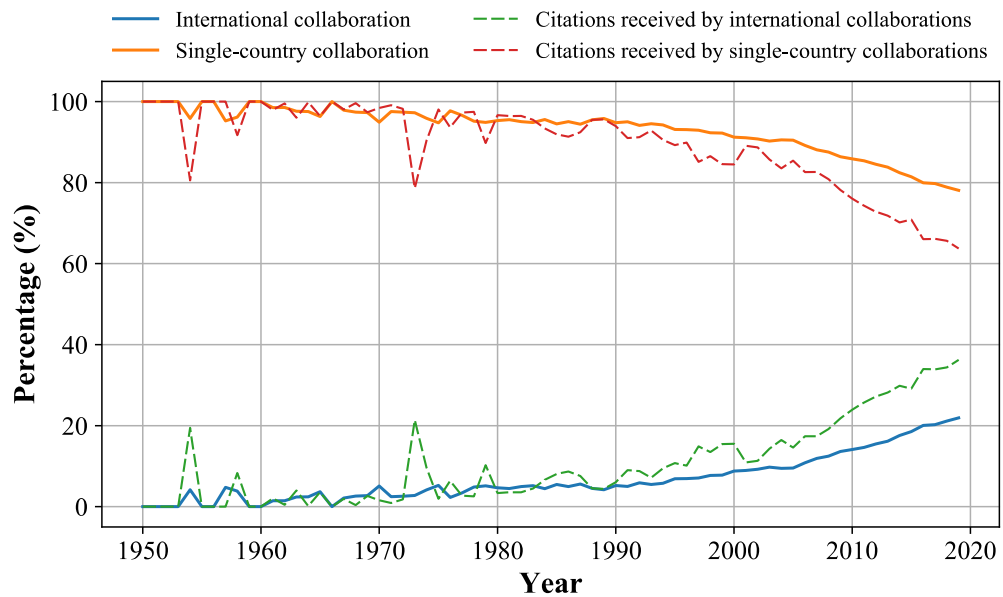

Figure 3. The changes in the annual percentage of nationally and internationally collaborated AI papers, as well as the percentage of citations received by nationally and internationally collaborated AI papers, in the period of 1950-2019.

To further investigate the international collaboration in AI across different countries, we choose the top 30 high-productive countries for analysis. The number of AI papers accounting for around 95% of the whole AI papers. Figure 4 shows the international collaboration in AI, from which we see that there are only 18 countries whose ratio of international collaboration is higher than 30% (the yellow dotted line). The top 3 highly collaborative countries in AI are the United States, China, and Japan, however, of which the ratios of international collaboration are quite low, i.e., 25% for the United States, 19.5% for China and 16.7% for Japan. As expected, the ratios of the European countries that are geographically closer to each other is relative higher, such as the United Kingdom (38. 7%) and Germany (31.3%).

Regardless of the low ratio, the number of international collaborations in AI is significant. For example, the United States have the largest number of internationally collaborated AI papers (nearly 100,000). It is worth to note that the percentage of citations received by international collaboration (the red dotted line) has been always higher than the percentage of international collaboration (the yellow dotted line) in AI (Figure 4), indicating that the international collaborations may bring more citations. Moreover, the percentages of citations received by international collaborations were nearly twice as the percentage of international collaborations for India, Japan, Korea Rep., China, Poland and Czech Republic.

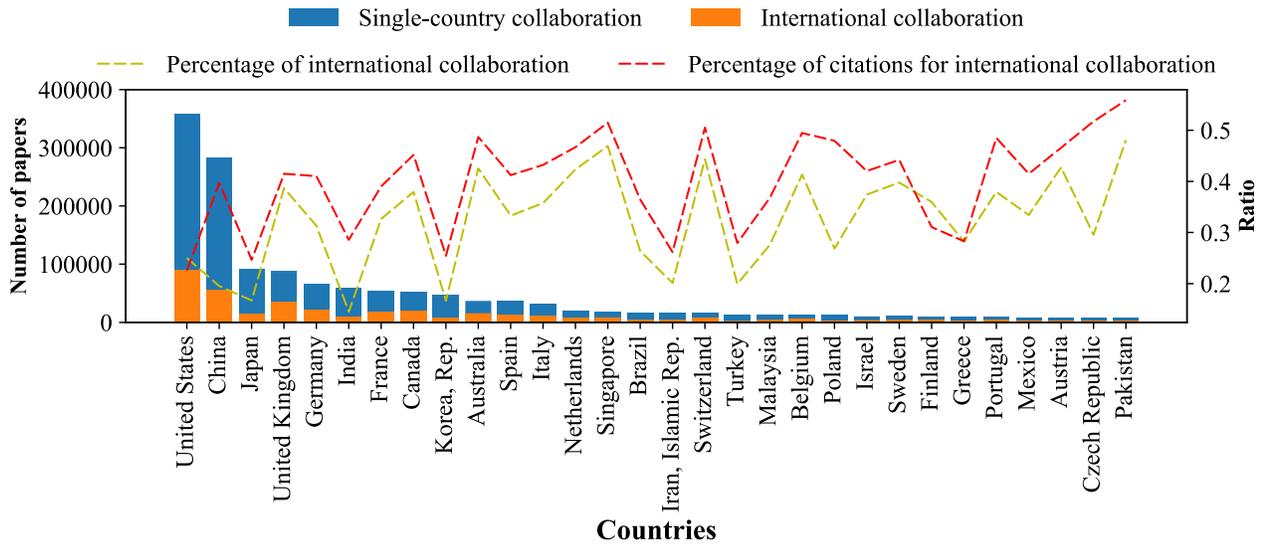

Figure 4. The number and percentage of national and international collaborations in the top 30 high productive countries in the field of AI during 1950 – 2019.

Figure 5 shows the results of social network analysis on the international collaborations between the top 30 high productive countries, from which we can see that the most internationally collaborative country in AI is the United States, followed by China, United Kingdom, Germany, Australia, Canada, and France, which are all developed countries except for China. Some developing countries (such as Brazil, Pakistan, and Iran) also had relatively high RDC values. Large part of their AI papers is internationally collaborated (Figure 4), though the academic production of them is low. Meanwhile, we can observe that the international collaborations between the top 30 countries were intense, especially for the collaborations between the United States and China, which indicated by the thickest edge in the network (Figure 5). Similar results can also be found in Appendix Information 1, in which we examined the importance of a country in global AI collaboration based on the between centrality and closeness centrality.

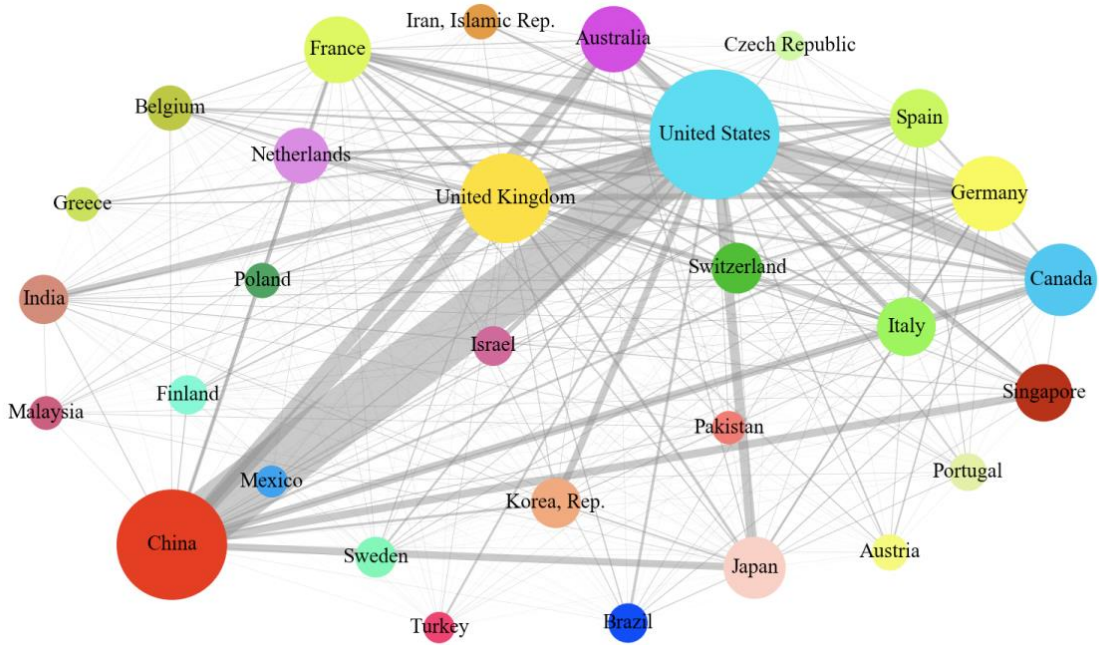

Figure 5. Social network analysis of the international collaboration among the top 30 high productive countries, the size of circles indicates the RDC of countries, and the width of the edges indicate the times of collaboration.

## 4.2. Trends of the spatial, economic, cultural, academic, and industrial separations of international collaboration in AI

We examined the evolution of spatial, economic, cultural, academic, and industrial separations of international collaboration in AI over the years, which are calculated by the annual average of geographic, economic, cultural, academic, and industrial distances between the affiliations of coauthors of articles, respectively. Figure 6 shows the changes in the five kinds of separations in international collaboration over the years, from which we can observe that all the five kinds of separations in international collaborations have increased, except for the cultural distance in masculinity/felinity dimension and the industrial distance. Specifically, the geographic distance has grown form 6,000 km in 1950 to 8,000 km in 2019 and the economic distance has increase from 0.3 in 1980 to 1.4 in 2019. Except for the cultural distance in masculinity/felinity dimension, the values of other four kinds of cultural distances have all increased by around 20 during the past 70 years. Meanwhile, we can see that the academic distances have shown a swift increase from 1950 to 2019, while the industrial distance have shown a decline during this period.

We further employed correlation analysis (including Pearson and Spearman correlation analysis) to explore the statistical significance of the trends of these separations of international collaboration in AI. As shown in Table 1, the increasing trends of all distances (except for the cultural distance in masculinity/felinity dimension and the industrial distance) were significant with the positive correlations with the years, especially for the academic distance in academic production ($r_{Pearson} = 0.890$, $p < 0.001$; $r_{Spearman} = 0.899$, $p < 0.001$) and the economic distance ($r_{Pearson} = 0.858$, $p < 0.001$;

$r_{Spearman} = 0.871$, p < 0.001). These findings indicate that researchers were seeking international collaborators different from themselves in various aspects. Note, they are more likely to collaborate with researchers from the countries sharing similar values on gender culture after 2000. Besides, we can also observe that the uncertainty avoidance distance (i.e., **dUA**) between collaborative countries showed a decline trend.

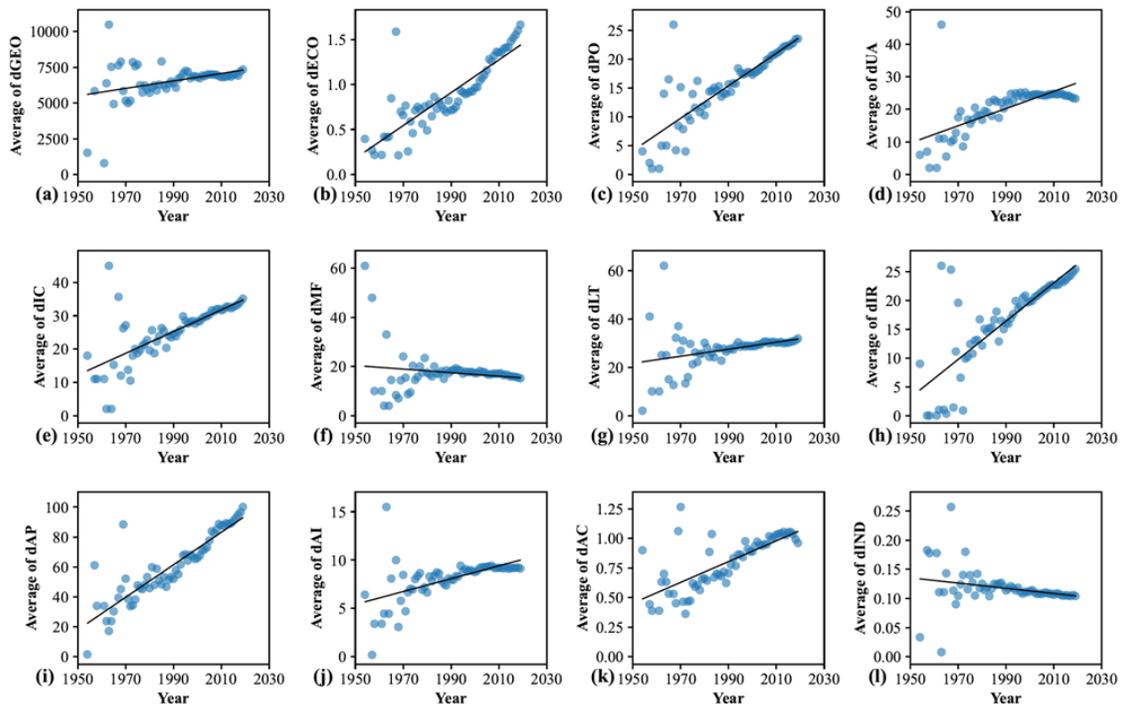

Figure 6. The changes in the geographic, economic and cultural separations in scientific collaboration over the years. (Note, **dGEO**: geographic distance; **dECO**: economic distance; **dPO**: cultural distance in power dimension; **dUA**: cultural distance in uncertainty avoidance dimension; **dIC**: cultural distance in individualism/collectivism dimension; **dMF**: cultural distance in masculinity/felinity dimension; **dLT**: cultural distance in the long-term orientation; and **dIR**: cultural distance in indulgence/restraint dimension; **dAP**: academic distance in academic production; **dAI**: academic distance in academic influence; **dAC**: academic distance in academic communication; **dIND**: industrial distance.)

Table 1. Correlation analysis of five kinds of distances between countries and years.

|  | Pearson | p | Spearman | p |
|---|---|---|---|---|
| dGEO | 0.354 | ** | 0.380 | ** |
| dECO | 0.858 | *** | 0.871 | *** |
| dPO | 0.840 | *** | 0.868 | *** |
| dUA | 0.664 | *** | 0.779 | *** |
| dIC | 0.717 | *** | 0.792 | *** |
| dMF | -0.160 |  | 0.046 |  |
| dLT | 0.332 | ** | 0.540 | *** |
| dIR | 0.801 | *** | 0.813 | *** |
| dAP | 0.890 | *** | 0.899 | *** |
| dAI | 0.563 | *** | 0.732 | *** |
| dAC | 0.729 | *** | 0.748 | *** |
| dIND | -0.257 | * | -0.495 | *** |

(Note, *: p<0.05; **: p <0.01; ***: p < 0.001.)

*4.3. Effects of the distance factors on the international collaboration in AI*

To investigate the relationships between the distance factors and the international collaboration in the field of AI, we plotted the distributions of probability of co-publications between countries over different distances. As shown in Figure 7, for all kinds of distances, the overall probability of international collaboration (co-publication) declines as the distance between countries increases (blue lines). We can also observe that these blue curves are not smooth. For example, in Figure 7(b), there is a little "hill" on the distribution of co-publication between countries when the economic distance ranges from 2.5 to 3.5. Similar phenomenon can be found in the Figure 7(e), (i) and (j). By rechecking the data and the Figure 5, we speculated that this phenomenon may be caused by the United Stated and China, which are the first and second most internationally collaborative countries in the field of AI. We also found that the peaks of hills are very close to the distances between these two countries. Further, we plotted the distributions of co-publications between countries in the field of AI without the United States (green lines) and without China (orange lines). Interestingly, some of the hills evidently become smaller or disappeared (Figure 7 (b), (c), (d), (e), (g), (h), (j), and (l)). These findings indicate that super countries like the United States and China may break the structures of international collaborations in the field of AI (Jiang et al., 2018; Parreira et al., 2017). Therefore, we added two explanatory indicators into the following regression analysis, that is, the United States as a collaborator (CoUS) and China as a collaborator (CoCN). These two variables are defined as dummy variables: CoUS = 1 (or CoCN = 1) if one of the two countries is the United States (or China); or, CoUS = 0 (or CoCN = 0).

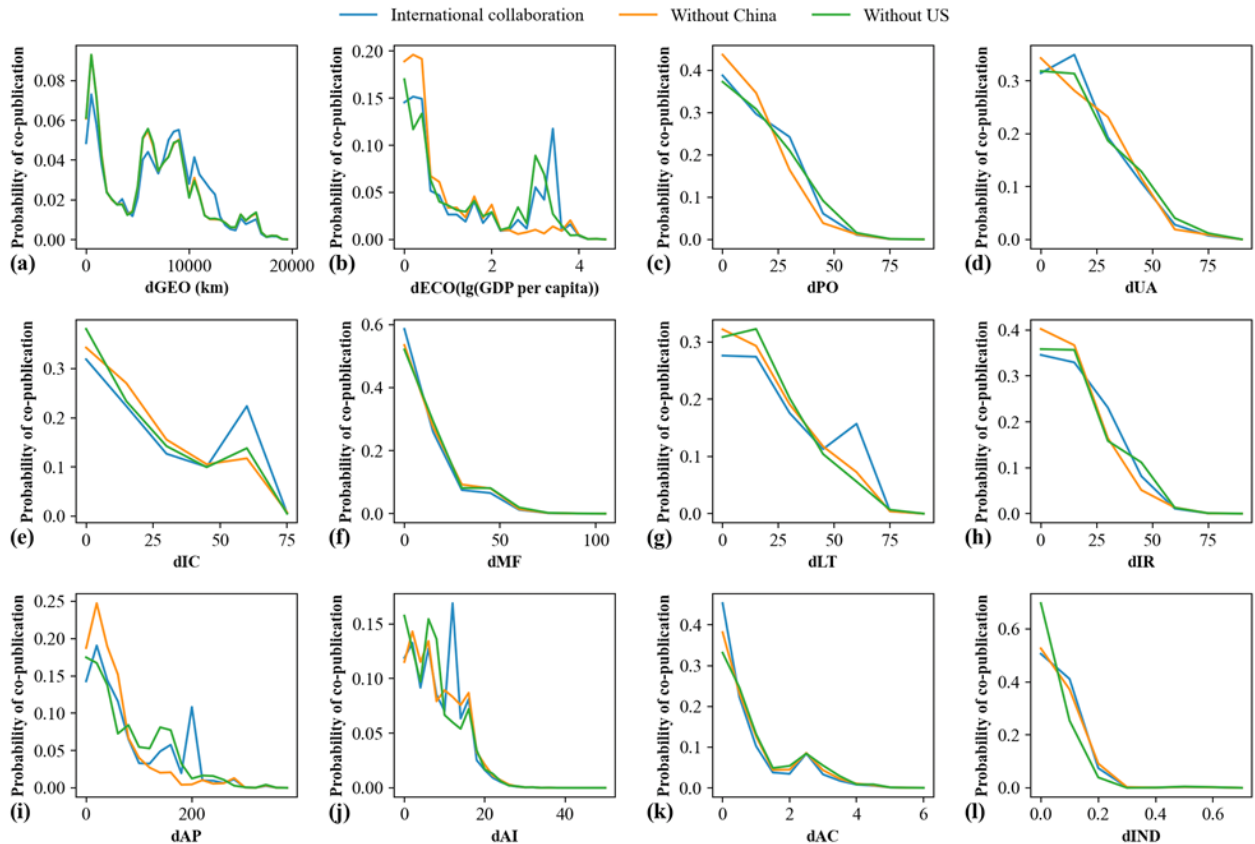

Figure 7. Effects of the distance factors on the probability of co-publication in the field of AI. (Note, **dGEO**: geographic distance; **dECO**: economic distance; **dPO**: cultural distance in power dimension; **dUA**: cultural distance in uncertainty avoidance dimension; **dIC**: cultural distance in individualism/collectivism dimension; **dMF**: cultural distance in masculinity/felinity dimension; **dLT**: cultural distance in the long-term orientation; and **dIR**: cultural distance in indulgence/restraint dimension; **dAP**: academic distance in academic production; **dAI**: academic distance in academic influence; **dAC**: academic distance in academic communication; **dIND**: industrial distance.)

We further examined the determinants of international collaboration in the field of AI using the linear regression model (OLSs) and the Zero inflated beta (ZIBeta) regression model. The result of the multicollinearity diagnosis of independent variables is shown in Appendix Information 2, from which we can see that all variables' variance inflation factors (VIFs) are lower than 3, indicating that here is no significant multicollinearity in the data. As shown in Table 2, according to the results of AIC and Pseudo $R^2$, the ZIBeta regression model (AIC = -20,400.93 and Pseudo $R^2$ = 0.132) performed better than the linear regression model (AIC = -2,978.268 and Pseudo $R^2$ = 0.104). The coefficients of the ZIBeta model indicated several important findings. First, the geographic distance (dGEO), economic distance (dECO), and academic distances (dAP, dAI and dAC) were negatively related to the international collaboration between countries, especially for the dECO (coefficient = -1.618, $p < 0.001$). This finding indicates that countries, who are geographically closer or who have similar economic and academic characteristics, tend to collaborate with each other in the field of AI. Second, the language factor (ENG) had a significant positive relationship with the international collaboration (coefficient = 0.127, $p < 0.001$), illustrating that courtiers with English as their official language tend to collaborate with countries with the same official language in the field of AI. Meanwhile, we can also find that the industrial distance also showed an obvious positive relationship with the international collaboration between countries. Third, CoUS (coefficient = 0.513, p < 0.001) and CoCN (coefficient = 0.617, p < 0.001) were also significant positively related with the international collaboration in AI, which indicate that the participation of the United States and China have promoted the international collaboration in the field of AI. In addition, although there were no statistically significant relationships between the cultural distances (dPO, dUA, dIC, dMF, dLT and dIR) and the international collaboration in AI (Figure 7 (c) – (h) and Table 2).

**Table 2 Effect of distance indicators on international collaboration**

|  | OLS | | ZIBeta | |
| --- | --- | --- | --- | --- |
|  | coefficients | p-value | coefficients | p-value |
| dGEO | -0.142(***) | <2e-16 | -0.399(***) | <2e-16 |
| dECO | -0.447(***) | <2e-16 | -1.618(***) | 2.80e-8 |
| dPO | 0.015 | 0.302 | -0.150 | 0.146 |
| dUA | 0.077 | 0.196 | 0.272 | 0.070 |
| dIC | 0.081 | 0.137 | 0.083 | 0.379 |
| dMF | 0.021 | 0.198 | 0.076 | 0.488 |
| dLT | -0.031 | 0.06 | -0.350 | 0.506 |
| dIR | -0.025 | 0.048 | -0.097 | 0.341 |
| ENG | 0.008(**) | 0.03 | 0.127(***) | 5.80e-6 |
| dAP | -0.163(***) | <2e-16 | -0.975(***) | <2e-16 |
| dAI | -0.144(***) | 1.68e-13 | -0.720(**) | 0.001 |
| dAC | -0.515(***) | 0.000 | -0.265(***) | 5.23e-6 |
| dIND | 0.484(***) | <2e-16 | 1.357(*) | 0.013 |
| CoUS | 0.035(***) | <2e-16 | 0.513(***) | <2e-16 |
| CoCN | 0.131(***) | <2e-16 | 0.617(***) | <2e-16 |
| Pseudo $R^2$ | 0.104 |  | 0.132 |  |
| AIC | -2,978.268 |  | -20,400.93 |  |

(Note, *: p< 0.05; **: p < 0.01; ***: p < 0.001. **dGEO**: geographic distance; **dECO**: economic distance; **dPO**: cultural distance in power dimension; **dUA**: cultural distance in uncertainty avoidance dimension; **dIC**: cultural distance in individualism/collectivism dimension; **dMF**: cultural distance in masculinity/felinity dimension; **dLT**: cultural distance in the long-term orientation; and **dIR**: cultural distance in indulgence/restraint dimension; **dAP**: academic distance in academic production; **dAI**: academic distance in academic influence; **dAC**: academic distance in academic communication; **dIND**: industrial distance.)

## 5. Discussion and conclusions

Artificial intelligence (AI) has emerged as an important part of modern science, with strong theoretical foundations and extensive methodological progress (Frank, Wang, et al., 2019; Perc et al., 2019). With more complicated and integrated issues faced by humans, international collaboration has become imperative in the field of AI (Luengo-Oroz et al., 2020, p. 19). However, efficient and stable collaboration among researchers from different countries was still a challenge because of distance factors between countries, such as geographic distance (Pan et al., 2012) and economic distance (Jiang et al., 2018). Therefore, it is significant to examine the evolving trends of international collaboration in the dimensions of these factors in the field of AI. It is also important to understand how these distance factors have influenced the international collaboration in the field of AI.

Our results showed that international collaboration is still not prevalent in the field of AI, accounting for only 15.7% of all collaborative AI papers from 1950 to 2019. Although the AI papers were authored by researchers from almost all the countries worldwide, there are only 18 countries whose ratio of papers with international authors is more than 30%. The top two most highly productive countries (i.e., United States and China) have also played important role in collaborations among countries in the field of AI by the value of RDC (Figure 4 and 5). The number of international collaborations in the field of AI have increased over years although the low ratios in different countries. Meanwhile, almost all the collaboration separations between countries have also shown a significantly increasing trend over years in different dimensions (Figure 6 and Table 1). We further found that several distance factors have influenced the international collaboration between countries in the field of AI, such as the geographic distance, economic distance, and industrial distance.

The negative relationship between the economic distance and the international collaboration in the field of AI was impressive. In fact, several studies have demonstrated the significance of economic factors, such as the gross domestic product (GDP), in scientific collaboration and communication between countries in the world (Choi et al., 2015; Parreira et al., 2017). Overall, countries with high GDP (e.g., United States, China and Japan) have stood out in scientific investment, contributing an increase in scientific activities and outputs. We discovered that highly collaborative country pairs in the field of AI were those have similar economic strength (GDP), which is consistence with previous researches and indicates that economic similarities have promoted international collaborations (Fernández et al., 2016; Gui et al., 2019; Jiang et al., 2018). Actually, the economic agreements between those countries not only have promoted international trade, but also have provided the resources and opportunities for scientific mobility, scholar exchange and international collaboration (Acosta et al., 2011; Fernández et al., 2016).

Similar with the economic distance, we found that highly collaborative countries in the field of AI had the similar academic characteristics. That is, the academic distances between countries were negatively related to the international collaboration in the field of AI. Several studies have concluded that, at the author level, especially in authors with high academic production and high academic influence tend to collaborate with others who have the similar academic background, due to their homophily preferences (C. Zhang et al., 2018). Our findings indicate that this collaboration pattern could also be applied to international collaboration at the country level. Besides, those countries with the similar academic communication may tend to communicate with each other more frequently, which expand the possibility of collaboration.

The international collaboration in the field of AI has also been influenced by the industrial distance. The industrial involvement with significant investment and powerful computational resource, which has attracted AI talents and teams worldwide, has greatly contributed to the international collaboration in the field of AI. For example, companies like Google, Baidu and Facebook have become the leaders of the modern research of AI (Frank, Wang, et al., 2019; Wu et al., 2020). Our finding indicates that countries with similar industry-involvement in AI research tend to have lower number of collaborative AI papers. The complementary resources between companies and universities among these countries, have provided scholars and developers an international and advanced collaborative platform, which has greatly accelerated the pace of AI innovations.

Surprisingly, the geographic distance still has negative influence on the international collaboration in the field of AI, that is, geographically distant countries tend to collaborate with each other less than closer countries. This finding indicates that the function of information and communication technology (ICT) seems to be overestimated. Even for the field of AI, a discipline that closely depend on the information technology and mainly complete the experiments in virtual environments, the geographical distance still has a statistical significance in the international collaboration, although the coefficient is not high.

In addition, the results showed that the participation of the United States (or China) has a significant positive relationship with the international collaboration in the field of AI. The two countries stand out in the economic strength, academic production, funding support, industrial strength (computational resources) and international AI talents, providing enough resources and opportunities for researchers to participate in international programs (Börner et al., 2020). Moreover, the Matthew effect might have enhanced these advantages. Besides, we found that the language factor (i.e., English as a primary or official language) had shown a significant influence on the international collaboration in the field of AI, indicating that linguistic barrier could be one of the factors hindering the international collaboration between countries. This could be interpreted by that researchers from countries with English as their primary or official language can better exchange their ideas and understand the embedded meanings during their communications.

The current study has several limitations. First, we only consider the influence of distance factors between countries on the international collaboration in the field of AI, and we didn't examine the influence of other potential factors (e.g., the leadership of authors and the research topic diversity in different countries) on the international collaboration in the field of AI. Our future studies will take into account these factors to offer more insight. Second, the data used in our study is limited to the AI articles harvested from the Microsoft Academic Graph (MAG). Some other data source on artificial intelligence (such as patents, and codes on the GitHub), in which international collaboration related information were recorded, should be included in the future work. Third, we didn't explore the causal relationships between the international collaboration of AI and the distance factors. Finally, although we have limited our analysis into the field of AI, the analysis framework proposed in our paper could also be extended to examine how these distance factors influence the international collaboration in other fields. The extended studies in other disciplines will provide us a deeper and more comprehensive understanding of the relations between distance factors and international collaboration. In addition, we also intend to examine the underlying causal relationships between distance factors and international collaboration.

## Acknowledgement

This work was supported by the National Science Foundation of China (71420107026, 91646206). The supported by the Wuhan University (student exchange program) during a visit by Xuli Tang to the University of Texas at Austin is acknowledged. The support provided by the China Scholarship Council (CSC) during a visit by Xin Li to the University of Texas at Austin is acknowledged (No. 201806270047). The author would like to express special gratitude to Prof. Ying Ding for her valuable comments and editorial assistance.

## *Appendix Information 1*

**(1) Betweenness centrality:**

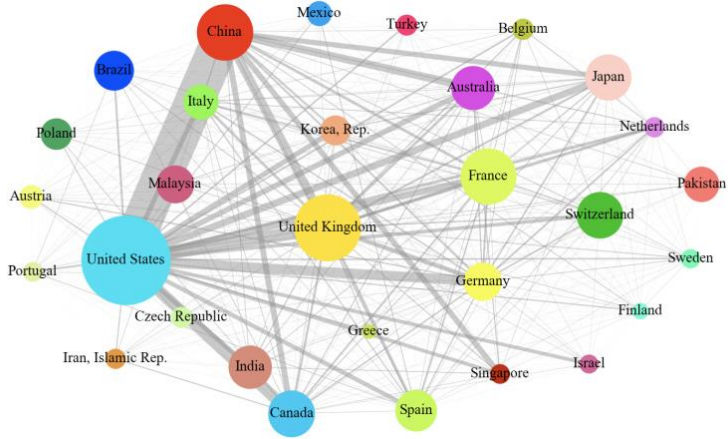

**Ranking:** United States > United Kingdom > China > France > Canada > Switzerland > Japan > Australia

**(2) Closeness centrality:**

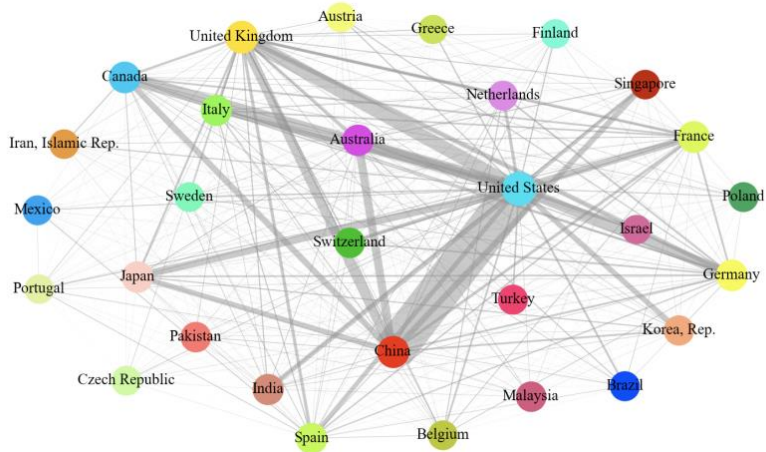

**Ranking:** United States > United Kingdom > China > France > Canada > Japan > Germany > Australia

*Appendix Information 2*

Table A1. Variance inflation factors (VIFs) and Pearson's correlation for the variables.

| | OLS VIF | ZIBeta VIF | $DIC_{ij}$ | dGEO | dECO | dPO | dUA | dIC | dMF | dLT | dIR | ENG | dAP | dAI | dAC | dIND | CoUS | CoCN |
|---|---|---|---|---|---|---|---|---|---|---|---|---|---|---|---|---|---|---|
| $DIC_{ij}$ | -- | -- | 1 | | | | | | | | | | | | | | | |
| dGEO | 1.12 | 1.23 | -0.16*** | 1 | | | | | | | | | | | | | | |
| dECO | 1.40 | 1.44 | -0.16*** | 0.05*** | 1 | | | | | | | | | | | | | |
| dPO | 1.51 | 1.44 | -0.01 | 0.03* | 0.18*** | 1 | | | | | | | | | | | | |
| dUA | 1.09 | 1.07 | 0.03 | 0.12*** | -0.01 | 0.06*** | 1 | | | | | | | | | | | |
| dIC | 1.69 | 1.70 | -0.02 | 0.15*** | 0.34*** | 0.49*** | 0.03* | 1 | | | | | | | | | | |
| dMF | 1.26 | 1.16 | 0.00 | -0.04** | -0.01 | 0.33*** | 0.04** | 0.20*** | 1 | | | | | | | | | |
| dLT | 1.20 | 1.17 | -0.05*** | 0.06*** | 0.04** | 0.06*** | 0.08*** | 0.11*** | 0.14*** | 1 | | | | | | | | |
| dIR | 1.11 | 1.10 | -0.04** | 0.22*** | -0.01 | 0.15*** | 0.02 | 0.15*** | 0.10*** | 0.12*** | 1 | | | | | | | |
| ENG | 1.17 | 1.24 | 0.01 | 0.12*** | 0.14*** | -0.03* | -0.04** | 0.04** | -0.25*** | -0.19*** | -0.05*** | 1 | | | | | | |
| dAP | 1.75 | 1.53 | -0.07*** | 0.07*** | 0.24*** | 0.01 | 0.22*** | 0.14*** | -0.04** | 0.21*** | -0.08*** | 0.02 | 1 | | | | | |
| dAI | 1.50 | 1.75 | -0.09*** | 0.05*** | 0.41*** | 0.29*** | 0.01 | 0.39*** | 0.14*** | -0.00 | 0.07*** | 0.05*** | 0.20*** | 1 | | | | |
| dAC | 1.38 | 1.14 | -0.06*** | -0.07*** | 0.10*** | 0.07*** | 0.14*** | 0.10*** | 0.08*** | 0.06*** | -0.05*** | -0.12*** | 0.42*** | 0.11*** | 1 | | | |
| dIND | 1.14 | 1.18 | 0.07*** | 0.02 | 0.17*** | 0.06*** | -0.03* | 0.11*** | 0.04** | 0.14*** | -0.04** | 0.05*** | 0.08*** | 0.13*** | -0.06*** | 1 | | |
| CoUS | 1.21 | 1.51 | 0.12*** | 0.10*** | 0.04** | 0.05*** | 0.00 | 0.22*** | -0.01 | -0.05*** | 0.02 | 0.14*** | -0.03* | 0.27*** | -0.03* | 0.22*** | 1 | |
| CoCN | 1.21 | 1.41 | 0.06*** | 0.08*** | 0.00 | -0.03* | 0.12*** | -0.02 | 0.01 | 0.17*** | 0.00 | -0.07*** | 0.32*** | -0.06*** | -0.02 | -0.01 | -0.01 | 1 |

*Appendix Information 3*

Considering the economy has grown rapidly, we estimated the model in four sub-periods (He et al., 2020; Tang et al., 2020). Details on AI publication in each period is shown in Table A2.

Table A2. Information on AI publications in the four sub-periods

|  | Time range | No. of co-publications | No. of international co-publications | Percentage |
| --- | --- | --- | --- | --- |
| Embryo | 1950-1999 | 130,741 | 8,378 | 6% |
| Stable | 2000-2007 | 236,685 | 24,818 | 10% |
| Machine Learning | 2008-2013 | 399,809 | 60,110 | 15% |
| Deep learning | 2014-2019 | 527,409 | 109,493 | 21% |
| Total | 1950-2019 | 1,294,644 | 202,799 | 16% |

Table A3 reports the results of the effects of distance indicators on the international collaboration of AI based on panel data. The models based on panel data have the consistent results with the original model.

Table A3. Effect of distance indicators on international collaboration of AI based on panel data

| Explanatory variables | OLS (Dependent variable *100) | | | | ZIBeta | | | |
| --- | --- | --- | --- | --- | --- | --- | --- | --- |
|  | Model 1 (1950-1999) | Model 2 (2000-2007) | Model 3 (2008-2013) | Model 4 (2014-2019) | Model 5 (1950-1999) | Model 6 (2000-2007) | Model 7 (2008-2013) | Model 8 (2014-2019) |
| dGEO | -0.035(*) | -0.069(***) | -0.126(***) | -0.191(***) | -0.386(*) | -0.696(***) | -0.650(***) | -0.849(***) |
| dECO | -0.078(***) | -0.091(***) | -0.159(***) | -0.237(***) | -0.535(*) | -0.952(***) | -1.027(***) | -1.290(***) |
| dPO | 0.022 | 0.010 | 0.003 | 0.021 | -0.028 | -0.0269 | -0.0254 | -0.0243 |
| dUA | 0.005 | 0.047 | 0.084 | 0.089 | -0.042 | -0.155 | 0.162 | 0.240 |
| dIC | -0.024 | 0.042 | 0.067 | 0.095 | -0.461 | -0.101 | 0.021 | 0.112 |
| dMF | 0.052 | 0.017 | 0.016 | 0.010 | 0.061 | 0.068 | 0.034 | 0.139 |
| dLT | -0.042 | -0.017 | -0.021 | -0.059 | -0.315 | -0.192 | -0.281 | -0.392 |
| dIR | -0.035 | -0.004 | -0.028 | -0.018 | -0.011 | 0.236 | -0.110 | -0.091 |
| ENG | 0.018 | 0.002 | 0.009(***) | 0.025(***) | 0.362(**) | 0.394(***) | 0.203(**) | 0.299(***) |
| dAP | 0.052 | -0.001 | -0.183(***) | -0.517(***) | -0.861(***) | -0.773(***) | -0.928(***) | -0.818(***) |
| dAI | -0.023 | -0.091(***) | -0.055(***) | -0.164(***) | -0.441 | -0.526(*) | -0.528(**) | -0.697(***) |
| dAC | -0.025 | -0.008 | -0.033(**) | -0.026 | -0.374 | -0.771(***) | -0.435(***) | -0.217(*) |
| dIND | -0.006 | 0.094(***) | 0.099(***) | 0.165(***) | -0.408 | -0.132 | 0.103 | 0.177(*) |
| CoUS | 0.031 | 0.121(***) | 0.317(***) | 0.602(***) | 0.915(***) | 0.974(***) | 0.964(***) | 0.205(***) |
| CoCN | 0.034 | 0.034(*) | 0.222(***) | 0.592(***) | 0.200 | 0.789(***) | 0.458(***) | 0.840(***) |
| Pseudo $R^2$ | 0.016 | 0.071 | 0.093 | 0.116 | 0.052 | 0.107 | 0.115 | 0.139 |